\documentclass[11pt]{article}

\usepackage{
graphicx,amsmath,amssymb,bm}

\newtheorem{prop}{Prop}[section]
\newtheorem{lemma}[prop]{Lemma}

\newtheorem{theorem}[prop]{Theorem}

\newtheorem{corollary}[prop]{Corollary}

\title{
Absence of replica symmetry breaking  \\
in the transverse and  longitudinal \\ random field Ising model 
}
\author{C. Itoi \\ 
Department of Physics, GS $\&$ CST, Nihon University, \\
Kanda-Surugadai, Chiyoda, Tokyo 101-8308, Japan} 
\begin{document}
\maketitle


\maketitle
\begin{abstract}
It is  proved that replica symmetry is not broken
in  the  transverse and  longitudinal random field Ising model. 
In this model, 
the variance of spin overlap of any component  
vanishes in any dimension almost everywhere in the coupling constant space
 in the infinite volume limit.  The weak Fortuin-Kasteleyn-Ginibre property in this model and the
Ghirlanda-Guerra identities in artificial  models
in a path integral representation based on the Lie-Trotter-Suzuki  formula
enable us to extend  Chatterjee's proof for the random field Ising model 
to the quantum model. 
\end{abstract}
\section{Introduction}
Replica symmetry breaking  is known to be  a non-trivial phenomenon in systems with quenched disorder.
This phenomenon  in mean field spin glass models has been studied deeply, since Talagrand proved  the Parisi conjecture \cite{Pr} for the Sherrington-Kirkpatrick (SK) model \cite{SK} in a mathematically rigorous manner \cite{T2}. 
When replica symmetry is broken,  the observed value of an observable in a typical sample 
differs from its sample expectation  with finite probability,
even though all samples in the sample ensemble are synthesized using exactly the same method. 
Theoretical physicists and mathematicians have been  seeking this phenomenon also in more realistic  short range spin glass models, such as the Edwards-Anderson (EA) model \cite{EA}, however, only a few rigorous results for the replica symmetry breaking
have been obtained in low temperature region in short range systems.
Nishimori and Sherrington showed that the replica symmetry breaking does not
occur on the Nishimori line located out of the spin glass phase in the EA model \cite{NS1,N}.  
Recently, Chatterjee proved a remarkable theorem that replica symmetry is not broken in the Ising model with a longitudinal 
Gaussian random field in any dimension
almost everywhere in the coupling constant space \cite{C2}.  It was shown that the variance of overlap vanishes in the system with  
the Fortuin-Kasteleyn-Ginibre (FKG) property using the Ghirlanda-Guerra identities. 
In the present paper, we extend his argument to quantum systems with the weak FKG property.
This is a first rigorous result for replica symmetry breaking in quantum disordered systems with short range interactions.

\section{Definitions and main result}

We study disordered quantum spin systems on $d$-dimensional cubic lattice
$V_L:= [1,L]^d \cap {\mathbb Z}^d$  and their corresponding 
classical spin systems on $(d+1)$-dimensional cubic lattice $W_{L,M} =V_L \times  T_M$, where
$ T_M:= [1,M] \cap {\mathbb Z}$ with positive integers $L$ and $M$. Let $B_L$ be a collection of interaction bonds which are
translations of a pair of sites in $V_L$. One of the most important example is given by nearest neighbor bonds
$B_L=\{\{ x,y\}|x,y \in V_L, |x-y|=1 \}$.
A spin operator  $S^{i}_x$ $(i=1,2,3)$ at  a site $x \in V_L$
on a Hilbert space ${\cal H} :=\bigotimes_{x \in V_L} {\cal H}_x$ is
defined by a tensor product of the Pauli matrix $\frac{1}{2}\sigma^i$ acting on ${\cal H}_x \simeq {\mathbb C}^{2}$ and unities.
These operators are self-adjoint and satisfies the commutation relation
$$
[S_x^1,S_y^2]=i \delta_{x,y} S_x^3 ,  \ \  \ \ \  [S_x^2,S_y^3]=i \delta_{x,y}  S_x^1, \ \ \  \ \ 
[S_x^3,S_y^1]=i \delta_{x,y} S_x^2,
$$
and the spin at each site $x$ has a fixed magnitude 
$$
\sum_{j=1}^3 (S_x^j)^2 =\frac{3}{4} {\bf 1}.
$$
We study the following Hamiltonian
\begin{equation}
H_V(S, g):= A(S^1,g^1)+B(S^3,g^3), 
\label{hamil}
\end{equation}
consisting of non-commuting  two terms $A$ and $B$ defined by
\begin{eqnarray}
&&A(S^1,g^1):=-\sum_{x \in V_L} J_1 g_x^1 S_x^1, \\
&&B(S^3, g^3) :=-\sum_{\{x,y \}\in B_L} S_x^3 S_y^3-\sum_{x \in V_L} ( J_3 g_x^3 +c)S_x^3,  
\end{eqnarray}
where  $(g_x^i)_{x \in V_L,i=1,3}$ are standard Gaussian i.i.d. random variables and
$J_1, J_3, c \in {\mathbb R} $ are coupling constants.

Here, we define Gibbs state for the Hamiltonian.
For a positive $\beta $,  the  partition function is defined by
\begin{equation}
Z_V(J, g) := {\rm Tr} e^{ - \beta H_V(S,g)}\end{equation}
where the trace is taken over the Hilbert space ${\cal H}$.

Let  $f$ be an arbitrary function 
of spin operators $S_x^i,  (x \in V_L, i=1,2,3)$.  
The  expectation of $f$ in the Gibbs state is given by
\begin{equation}
\langle f(S^i) \rangle=\frac{1}{Z_V(J,g)}{\rm Tr} f(S^i)  e^{ - \beta H_V(S,g)}.
\end{equation}
Here, we introduce a fictitious time  $t \in [0,1]$ and define a time evolution of operators with the Hamiltonian.
Let $O$ be an arbitrary self-adjoint operator, and we define an operator valued function  $O(t)$ of $t\in[0,1]$  by
\begin{equation}
  O(t):= e^{-tH} O  e^{tH}.
\end{equation}
Furthermore, we define the  Duhamel expectation of time dependent operators 
$ O_1(t_1),  \cdots,  O_k(t_k)$  by
$$
( O_1,  O_2,\cdots,   O_k)_{\rm D} :=\int_{[0, 1]^k} dt_1\cdots dt_k \langle {\rm T}[ O_1(t_1)  O_2(t_2) \cdots  O_k(t_k) ]\rangle,
$$
where the symbol ${\rm T}$ is a multilinear mapping of the chronological ordering.
If we define a partition function with arbitrary self adjoint operators  $O_0, O_1, \cdots, O_k$ and real
numbers $x_1, \cdots, x_k$
$$
Z(x_1,\cdots, x_k) := {\rm Tr} \exp \beta \left[O_0+\sum_{i=1} ^k x_i O_i \right],
$$
the Duhamel expectation of $k$ operators represents
 the $k$-th order derivative of the partition function 
 \cite{Cr,GUW,S}
$$\beta^k( O_1,\cdots,  O_k)_{\rm D}=\frac{1}{Z}
\frac{\partial ^k Z}{\partial x_1 \cdots \partial  x_k}.
$$

To study replica symmetry breaking, we consider $n$ replicated spin model defined by the following
Hamiltonian
\begin{equation}
\sum_{\alpha=1} ^n H_V(S^\alpha,g).
\label{hamilrep}
\end{equation}
The overlap operator $R^i_{\alpha,\beta} (i=1,2,3)$ between different replicated spins  is defined by
$$R^i_{\alpha,\beta}=\frac{1}{|V_L|} \sum_{x \in V_L}S^{i, \alpha}_x S^{i,\beta}_x,$$
for $\alpha,\beta = 1,2 \cdots, n$  and $\alpha\neq \beta$.

It is well-known that quantum spin systems on a $d$-dimensional lattice  can be represented as 
$(d+1)$-dimensional classical Ising systems \cite{Suzuki}.
The Lie-Trotter-Suzuki formula for the Hamiltonian (\ref{hamil}) 
$$
e^{-\beta  A-\beta B} = \lim_{M\rightarrow \infty} (e^{-\beta  A/M }e^{-\beta  B/M })^M
$$
and inserting $M$ resolutions of unity in eigenstates of $2S_x^3$ on ${\cal H}$
\begin{equation}
 {\bf 1} =\sum_{\sigma \in {\cal S}_V}{|\sigma \rangle}_{3~3}\langle \sigma |,
\label{unity}
\end{equation}
where  we define 
$$
S^i_x |\sigma \rangle_i = \frac{\sigma_x}{2} |\sigma \rangle_i.
$$
${\cal S}_V := \{-1,1\}^{V_L}$ is a set of eigenvalue configurations, 
 enable us to represent
the $d$-dimensional quantum spin system in the following  $(d+1)$-dimensional classical spin system
\begin{equation}
Z_V(J,g)=\lim_{M\rightarrow \infty} C_W \sum_{\sigma \in {\cal S}_W} e^{-\beta H_W(\sigma,g)},
\end{equation}
where the summation is taken over 
spin configurations ${\cal S}_W := \{ -1,1\}^{W_{L,M}}$ on the $(d+1)$-dimensional lattice $W_{L,M}$
 and the factor $C_W$  is independent of spin configurations.
In this representation, we impose the periodic boundary condition  on spin configuration $\sigma_{x,t+M}=\sigma_{x,t}$ with respect to 
$t \in T_M$ and free boundary condition with respect to $x \in V_L$.
For instance in the transverse field Ising model with longitudinal random field \cite{CK,CKP,Cr}, the Hamiltonian is given by
\begin{equation}
H_W(\sigma,g)= -\sum_{t \in T_M}\Big[ \frac{1}{4M} \sum_{\{x,y\}\in B_L} \sigma_{x,t}
\sigma_{y,t} +\frac{1}{2M} \sum_{x \in V_L}(J_3 g_x^3 +c)\sigma_{x,t} 
+\sum_{x\in V_L} K_x \sigma_{x,t} \sigma_{x,t+1}\Big],
\label{HamilW}
\end{equation}
where
$$
\tanh \beta K_x =e^{- \beta J_1 g_x^1 /M},
$$
and the factor is given by
\begin{equation}
C_W= \prod_{x\in V_L}  \Big| \frac{1}{2} \sinh \frac{\beta J_1 g_x^1}{M} \Big|^\frac{M}{2}.
\end{equation}

To obtain our main result,
we consider an artificial $(d+1)$-dimensional random field Ising  model with quenched i.i.d 
standard Gaussian random variables $(h^i_{x,t})_{(x,t) \in W_{L,M},i=1,3}$, 
and arbitrary  numbers $b_i, c \in {\mathbb R} $ for $i=1,3$. We define the following perturbed 
Hamiltonian
\begin{eqnarray}
H(b_1,b_3,c,h_1,h_3) := &&-\sum_{t \in T_M} ( \frac{1}{4M} \sum_{\{x,y\}\in B_L} \sigma_{x,t} \sigma_{y,t} +\sum_{x\in V_L} K_{x,t} \sigma_{x,t} \sigma_{x,t+1}) \nonumber \\
&&- \frac{1}{2M}\sum_{(x, t)\in W_{L,M}}(J_3g_x^3
+  b_3 \sqrt{  M} h_{x,t}^3+c )\sigma_{x,t}, \label{hamilu}
\end{eqnarray} 
such that $H(0,0,c,0,0)$  is identical to $H_{W}(\sigma,g)$ defined by (\ref{HamilW}). This model has operator representation 
\begin{equation}
Z_V(b) =  \lim_{M\rightarrow \infty} {\rm Tr} ~ \prod_{t=1} ^M (e^{\beta A_t/ M } e^{B_t/M} ),
\end{equation}
 where 
\begin{eqnarray}
&&A_t(S^1,g^1,h^1):=-\sum_{x \in V_L} (J_1 g_x^1  +b_1h^1_{x,t})S_x^1, \\
&&B_t(S^3, g^3,h^3) :=-\sum_{\{x,y \}\in B_L} S_x^3 S_y^3-\sum_{x \in V_L} ( J_3 g_x^3 +  b_3 \sqrt{M} h^3_{x,t}+c 
) S_x^3,  
\nonumber
\end{eqnarray} 

For lighter notation,  we denote 
\begin{equation}
H_1(b):=H(b,0,c,h,0), \ \ \ \ 
H_3(b):=H(0,b,c,0,h)
\label{hamilu2}
\end{equation}
and define  partition function $Z_i(b)$ by
$$
Z_i(b):=\sum_{\sigma \in {\cal S}_W } e^{-\beta H_i(b) },
$$
and define  functions $\psi_{i,L}$  and $p_{i,L}$ by
$$
\psi_{i,L} (b) :=\frac{1}{|V_L|} \log Z_i(b),  \ \ \ p_{i,L}(b) := {\mathbb E} \psi_{i,L} (b),
$$
where a sample expectation  ${\mathbb E}$  denotes expectation over all random fields  $(g_x^i)_{x\in V_L,i=1,3},$ 
$(h^i_{x,t})_{(x,t) \in W_{L,M},i=1,3}$.

Note that 
$$
Z_V=\lim_{M\rightarrow \infty}C_W Z_i(0),
$$
given by  (\ref{hamilu}) and (\ref{hamilu2}).
Hereafter, $\langle f \rangle_{b,i}$ denotes
the Gibbs expectation of a function $f: {\cal S}_W \rightarrow {\mathbb R}$
 with the Hamiltonian $H_i(b)$
 $$
 \langle f(\sigma) \rangle_{b,i}:= \frac{1}{Z_i(b)} \sum_{\sigma \in {\cal S}_W } f(\sigma) e^{-\beta H_i(b) },
 $$
 
Note that for $i=1,3$
$$
\langle f(2S^3) \rangle =\lim_{M\rightarrow \infty} \langle f(\sigma) \rangle_{0,i}
$$
in the representations (\ref{hamilu}) and (\ref{hamilu2}) for $H_i(b).$\\

In the present paper, we obtain the following main theorem for the transverse and longitudinal random field Ising model.

{\theorem \label{MT} 
Consider  the transverse and longitudinal random field Ising model defined by the Hamiltonian
(\ref{hamil}) and its replicated model (\ref{hamilrep}).
Almost everywhere in the coupling constant space, the infinite volume limit
$$\displaystyle{\lim_{L \rightarrow \infty }{\mathbb E}  \langle R^i_{1,2} \rangle},$$
 exists  for any $i = 1,2,3$ and 
 the variance of the overlap operator calculated in the replica symmetric Gibbs state vanishes
\begin{equation}
\lim_{L \rightarrow \infty }{\mathbb E}\langle ( R^i_{1,2} -{\mathbb E} \langle R^i_{1,2} \rangle )^2 \rangle=0.
\end{equation} 
 }\\
Theorem \ref{MT} shows that the overlap operator $R^i_{1,2}$ is self-averaging in this model. This implies that the observed value of the overlap 
operator $R_{1,2}^i$ converges in probability toward its Gibbs and sample expectation ${\mathbb E} \langle R^i_{1,2}\rangle $.  
Since the replica symmetric Gibbs expectation of the overlap operator is spin glass order parameter, the phase diagram should be 
unique if the sample is synthesized in the same method. 

There are two key techniques to prove Theorem \ref{MT}: the weak FKG property of the transverse and longitudinal random 
field Ising model  in the $(d+1)$-dimensional  representation and continuity of
an artificial perturbative $(d+1)$-dimensional   model with the  Ghirlanda-Guerra identities.  
Since a straightforward  extension of the  Ghirlanda-Guerra identities to quantum systems 
is not sufficient to judge  absence or appearance of replica symmetry breaking in quantum systems unlike the classical system \cite{C2}, 
we utilize the classical Ghirlanda-Guerra identities in artificial models given by Hamiltonians (\ref{hamilu}) and (\ref{hamilu2}). 
We prove that the expectation of the overlap operator is a continuous function of the perturbation parameter.
These results enable us to prove
Theorem \ref{MT} which shows absence of replica symmetry breaking in the transverse and longitudinal random field Ising model.

\section{Proof}

Here, we consider the perturbed model defined by the Hamiltonian (\ref{hamilu2}) in $d+1$ dimension.
For this model, there are  useful lemmas proved in the literature.
Here we present them as Lemma \ref{0}-\ref{FKGineq} without proofs. 
Lemma \ref{0} is proved as in \cite{AGL,CGP,CL,I}, Lemma \ref{freeva} and Lemma \ref{delta} are proved 
in \cite{I,I2}, and  Lemma \ref{FKGineq} is proved in \cite{FKG}.

{\lemma \label{0}
The following  infinite volume limit independent of boundary conditions  exists
$$
 p_{i}(b) = \lim_{L \rightarrow \infty}   \lim_{M \rightarrow \infty} {\mathbb E} \psi_{i,L}(b). 
$$
for each $(\beta, J_1,J_3,b,c) \in [0,\infty) \times {\mathbb R}^{4}$.}\\

{\lemma \label{freeva}  
For any $(\beta, J_1,J_3, b,c) \in [0,\infty) \times {\mathbb R}^{4}$,
there exists a positive number $C$ independent of $L$, such that  the  variance of 
$ \psi_L$ is bounded from the above  as follows
$$
\lim_{M\rightarrow\infty}{\mathbb  E} [\psi_{i,L}(b)- p_{i,L}(b) ]^2
 \leq   \frac{C  }{|V_L|}.
$$
}
Here,  we define two types of deviations of of an arbitrary operator $O$ by
$$\delta O:= O -\langle O \rangle, \ \ \  \Delta O := O- {\mathbb E} \langle O \rangle.$$
And,  also define two types of deviations an arbitrary function  $f:{\cal S}_W \rightarrow {\mathbb R}$ by
$$\delta f(\sigma):= f(\sigma) -\langle f(\sigma) \rangle_{b,i}, \ \ \  \Delta f(\sigma) := f(\sigma)- {\mathbb E} \langle f(\sigma) \rangle_{b,i}.
$$
Define an order parameter $m^i_L$ by
$$
m_L^i:=\frac{1}{|V_L|} \sum_{x\in V_L} S_x^i
$$
 and define the corresponding order parameter $\mu_L^i$ by
\begin{eqnarray}
&&\mu_L^1:= \frac{1}{4|V_L|M} \sum_{x\in V_L, t\in T_M}(1- \sigma_{x,t} \sigma_{x,t+1}), \\
&&\mu_L^3:= \frac{1}{2|V_L|M} \sum_{x\in V_L, t \in T_M} \sigma_{x,t}.
\end{eqnarray}

{\lemma  \label{delta}  
For any $(\beta, J_1,J_3, b,c) \in [0,\infty) \times {\mathbb R}^{4}$ with $\beta J_i\neq 0$,
there exists a positive number $C$ independent of $L$ , such that 
\begin{equation}
\lim_{M\rightarrow\infty} {\mathbb E}  \langle {\delta \mu^i_L } ^2\rangle_{b,i} \leq  \frac{C}{\beta J_i}\sqrt{\frac{1}{|V_L|}}. 
\label{a}
\end{equation}
}

Lemma \ref{delta} gives an upper bound of
Duhamel product for $b_i=0$ 
\begin{equation}
 {\mathbb E}  (  \delta m^i_L , \delta m^i_L )_{\rm D} \leq   \frac{C}{\beta J_i}\sqrt{\frac{1}{|V_L|}}. 
 \label{b}
\end{equation}

We say that the system satisfies the weak Fortuin-Kasteleyn-Ginibre (FKG) condition, if the one point function $\langle S_x^3\rangle$
is monotonically increasing function of $g_y^3$  at any  site $y \in V_L$ \cite{AGL}.  
The weak FKG condition is equivalent to the positive semi-definiteness of truncated Duhamel function
$$
(S_x^3,S^3_y)_{\rm D} -\langle S_x^3 \rangle \langle S_y^3 \rangle \geq 0,
$$
for any two sites $x,y \in V_L$. 
The $d$-dimensional transverse and longitudinal random field Ising model satisfies weak FKG condition, 
because of the  following lemma  for the corresponding $(d+1)$-dimensional classical model with positive semi-definite exchange interactions. 
To explain the FKG inequality, we define
a partial order $\leq$ over the set ${\cal S}_W$ of spin configurations and 
generalized monotonicity for a function of spin configurations. 
For two spin configurations $\sigma, \tau \in {\cal S}_W$,  we denote $\sigma \leq \tau$,
 if  $\sigma_x   \leq \tau_x $ for all $x \in W_{L,M}$. 
 We say that a function $f: {\cal S}_W \rightarrow {\mathbb R}$ is monotonically increasing in 
 a general sense, if $\sigma \leq \tau$ implies
 $f(\sigma) \leq f(\tau)$. The following FKG inequality  can be proved \cite{FKG}.
Note that the artificial Hamiltonians  $H_i(a,b)$ given by  (\ref{hamilu})  and (\ref{hamilu2}) satisfy the FKG condition as well.
Therefore, the one point function 
$
\langle  \sigma_{x,s} \rangle_{b,i}
$
is monotonically increasing function of  $h^3_{y,t}$.

{\lemma   \label{FKGineq} 
Let $f$ and $g$ be 
 monotonically increasing functions of spin configurations on $W_{L,M}$ in a general sense.
In the random field Ising model with positive semi definite exchange interactions, 
$f$ and $g$ satisfy the Fortuin-Kasteleyn-Ginibre inequality
\begin{equation}
\langle f(\sigma); g(\sigma)\rangle_{b,i}
 \geq 0,
 \end{equation}
  where a truncated correlation function is defined by
  \begin{equation}
\langle f(\sigma); g(\sigma)\rangle_{b,i}:=\langle f(\sigma)g(\sigma)\rangle_{b,i}- \langle f(\sigma) \rangle_{b,i} \langle g(\sigma)\rangle_{b,i}.
 \end{equation}
}

{\lemma \label{s1}  For arbitrary sites $w,x,y,z \in W_{L,M}$, 
\begin{eqnarray}
&&|\langle \sigma_x \sigma_y; \sigma_w \sigma_z  \rangle_{b,i}
 | 
\leq \langle (\sigma_x +\sigma_y); (\sigma_w +\sigma_z ) \rangle_{b,i}
\label{bondcorr} 
\end{eqnarray}
Proof.} Define  functions $f_{\pm}: {\cal S}_W\times W_{L,M}^2 \rightarrow {\mathbb R}$ by 
$$
f_{\pm} (\sigma, w,x):= (\sigma_w \pm 1)(1 \pm \sigma_x).
$$
For arbitrary fixed $w,x,y,z \in W_{L,M}$,  functions $f_+(,w,x)$  and $f_-(,y,z)$
of spin configurations are monotonically increasing in general sense.
From the FKG inequality,
\begin{eqnarray}
 &&\langle \sigma_x \sigma_y; \sigma_w \sigma_z  \rangle_{b,i}+\langle (\sigma_x +\sigma_y); (\sigma_w +\sigma_z ) \rangle_{b,i}
 \nonumber \\
&&=(\langle f_+(\sigma,w,x) ; f_+(\sigma,y,z)  \rangle_{b,i}
+\langle f_-(\sigma,w,x) ; f_-(\sigma, y,z)  \rangle_{b,i} )/2 \geq 0,
\end{eqnarray}
and also 
\begin{eqnarray}
&&-\langle \sigma_x \sigma_y; \sigma_w \sigma_z  \rangle_{b,i}+\langle (\sigma_x +\sigma_y); (\sigma_w +\sigma_z ) \rangle_{b,i},
 \nonumber \\
&&=(\langle f_+(\sigma,w,x) ; f_-(\sigma,y,z)  \rangle_{b,i}
+\langle f_-(\sigma,w,x) ; f_+(\sigma, y,z)  \rangle_{b,i} )/2 \geq 0.
\end{eqnarray}
 These inequalities give the  inequality (\ref{bondcorr}). 
 $\Box$\\

Next we evaluate Gibbs expectation of functions of  the overlap operators 
in the path integral representation with the Hamiltonian $H_i(bc)$. 
In these representations, we denote
\begin{eqnarray}
&&\rho^1_{\alpha,\beta} := \frac{1}{16|V_L| M } \sum_{(x,s) \in W_{L,M}} 
(1- \sigma^\alpha_{x,s} \sigma^\alpha_{x,s+1})(1-  \sigma^\beta_{x,s} \sigma^\beta_{x,s+1}), \nonumber \\
&&\rho^3_{\alpha,\beta} :=  \frac{1}{4|V_L|  M} \sum_{(x,s) \in W_{L,M}} 
\sigma^\alpha_{x,s} \sigma^\beta_{x,s}, \label{overlap}
\end{eqnarray}
for $\alpha \neq \beta$. Note that 
$$
\langle \rho^i_{\alpha,\beta} \rho^j_{\gamma,\delta} \rangle_{0,i} = (R^i_{\alpha,\beta}, R^j_{\gamma,\delta})_{\rm D},
$$
for $\alpha \neq \beta, \gamma \neq \delta$ and   for $i,j=1,3.$

{\lemma \label{MT0} In the model defined by the Hamiltonian (\ref{hamilu2}) 
with $J_1 \neq 0$ and $J_3 \neq 0$, the 
following expectations calculated in the replica symmetric Gibbs state
vanish

\begin{eqnarray}
&&\lim_{L \rightarrow \infty } \lim_{M \rightarrow \infty } {\mathbb E} [\langle {\rho^i_{1,2}}^2\rangle_{b,i} 
- \langle \rho^i_{1,2}\rangle_{b,i} ^2 ]=0, \label{1212} 
\end{eqnarray}
 for any $i = 1,3$. \\
 
 \noindent
 Proof.}  
 First, consider $i=3$
\begin{eqnarray}
&&{\mathbb E} [\langle{ \rho^3_{1,2} }^2\rangle_{b,3}  - \langle \rho^3_{1,2} \rangle_{b,3}^2 ] =
\frac{1}{16|V_L|^2M^2} \sum_{x,y \in W_{L,M}} {\mathbb E} [ \langle  \sigma_x \sigma_y \rangle_{b,3}^2-
 \langle  \sigma_x\rangle_{b,3}^2 \langle  \sigma_y\rangle_{b,3}^2]\nonumber \\
&&\leq  \frac{1}{16|V_L|^2M^2} \sum_{x,y \in V_L} {\mathbb E} | \langle  \sigma_x \sigma_y \rangle_{b,3}-
 \langle  \sigma_x\rangle_{b,3} \langle  \sigma_y\rangle_{b,3} || \langle  \sigma_x \sigma_y \rangle_{b,3}+
 \langle  \sigma_x\rangle_{b,3} \langle  \sigma_y\rangle_{b,3}|\nonumber \\
&& \leq \frac{1}{8|V_L|^2M^2} \sum_{x,y \in V_L}  {\mathbb E} 
  |   \langle  \sigma_x  ;\sigma_y \rangle_{b,3} |\nonumber \\
 &&= \frac{1}{8|V_L|^2M^2} \sum_{x,y \in V_L}  {\mathbb E} 
   \langle  \sigma_x  ;\sigma_y \rangle_{b,3}.
\end{eqnarray}
The final line is nonnegative because of the FKG inequality. Therefore,
 \begin{eqnarray}
  \lim_{M\rightarrow\infty}{\mathbb E}[\langle {\rho^3_{1,2}}^2\rangle_{b,3}  - \langle \rho^3_{1,2} \rangle_{b,3}^2 ]
  &\leq& 
 \lim_{M\rightarrow\infty} {\mathbb E}  \langle { \delta \mu^3_L}^2 \rangle_{b,3} 
 \leq  \frac{C}{\beta J_3\sqrt{[V_L|}},
  \end{eqnarray}
where we have used Lemma \ref{delta}.

For $i=1$,
 \begin{eqnarray}
 &&{\mathbb E} [\langle \rho^1_{1,2}\rho^1_{1,2}\rangle_{b,1}  - \langle \rho^1_{1,2} \rangle_{b,1}^2 ] 
  \leq  {\mathbb E}| \lim_{M\rightarrow \infty}\frac{1}{32|V_L|^2M^2}\sum_{x,y \in V_L}  \sum_{s,t \in T_M} {\mathbb E} |
  \langle  \sigma_{x,s}\sigma_{x,s+1}  ; \sigma_{y,t} \sigma_{y,t+1}\rangle_{b,1} | \nonumber \\
&&\leq   \lim_{M\rightarrow \infty}  \frac{1}{32|V_L|^2M^2} \sum_{x,y \in V_L}  \sum_{s,t \in T_M} {\mathbb E}
 |\langle   (\sigma_{x,s}+\sigma_{x,s+1} ) ; (\sigma_{y,t}+ \sigma_{y,t+1})\rangle_{b,1} |
 \nonumber \\
 &&\leq \lim_{M\rightarrow \infty}  \frac{1}{8|V_L|^2M^2} \sum_{x,y \in V_L}  \sum_{s,t \in T_M}
 | \langle  \sigma_{x,s}; \sigma_{y,t}\rangle_{b,1} |
 \nonumber \\
 &&= \lim_{M\rightarrow \infty}  \frac{1}{8|V_L|^2M^2} \sum_{x,y \in V_L}   \sum_{s,t \in T_M}{\mathbb E} 
 \langle  \sigma_{x,s}; \sigma_{y,t}\rangle_{b,1} 
 \nonumber \\
 &&=
 \lim_{M\rightarrow \infty} \frac{1}{2} {\mathbb E}  \langle { \delta \mu^3_L}^2 \rangle_{b,1}
   \leq \frac{C'}{\beta J_3\sqrt{[V_L|}}.
  \end{eqnarray}
 We have used the inequality (\ref{bondcorr}) in Lemma  \ref{s1}, Lemma \ref{delta}.
 These bounds give the limit. 
 $\Box$\\

In the original model with $b_1=b_3=0$, Lemma \ref{MT0} implies 
\begin{eqnarray}
&&\lim_{L \rightarrow \infty }{\mathbb E} [(R^i_{1,2},R^i_{1,2})_{\rm D}  - \langle R^i_{1,2}\rangle ^2 ]=0, \label{1212} 
\end{eqnarray}
 for any $i = 1,3$.


Here we regard  
\begin{equation}
\psi_{i,L}(h):= \frac{1}{| V_L|} \log Z_i(b)
\end{equation}  
as a function of disorder $h=(h_w)_{w  \in W_{L, M}}$.
Let $h$  and $({h}')$ be i.i.d. standard Gaussian random variables , and
define square root interpolating  random variables  with $ v \in [0,1]$ by
\begin{equation}
\sqrt{v} h_{w} +\sqrt{1-v} {{h}'_{w}},
\end{equation}
for $w\in W_{L,M}$.
Then, we define a generating  function $\gamma_i(v)$ of the parameter $v_i \in [0,1]$ by
\begin{equation}
\gamma_i(v) :=
 {\mathbb E} [{\mathbb E}' \psi_{i,L}(\sqrt{v} h+\sqrt{1-v} {h}' )]^2,
\end{equation}
where  
 ${\mathbb E }$ and  ${\mathbb E }'$ denote expectation
in $h$ and $h'$, respectively. This generating function $\gamma_i$ is  a generalization of a function introduced by Chatterjee \cite{C}.\\

{\lemma \label{1}  
For any $(\beta, J_1,J_3, b,c) \in [0,\infty) \times {\mathbb R}^{4}$, any positive integer
$L$, any positive integer $k$ and any $v_0 \in[0,1)$, 
an upper bound on the $k$-th order derivative of the function $\gamma_i$ is given by 
\begin {equation}
\frac{d^k \gamma_i}{{d v_i}^k} (v_0) \leq \frac{(k-1)! }{(1-v_0)^{k-1}}\frac{\beta^2 b^2}{4|V_L|}.
\end{equation}
For an arbitrary $v\in [0,1]$, the $k$-th order derivative of $\gamma_i$ is represented in the following
\begin{eqnarray}
\frac{d ^k \gamma_i}{{d v}^k}(v)  &=& \sum_{w_1 \in W_{L,M}}  \cdots \sum_{w_k \in W_{L,M}}
{\mathbb E} \left[{\mathbb E}' \psi_{i, L,w_1, \cdots, w_k}(\sqrt{v} h+\sqrt{1-v} {h}' )\right]^2.
\label{kth}
\end{eqnarray}
 Here we denote
$$
\psi_{i,L,w_1, \cdots, w_k}(h):=\frac{\partial^k \psi_{i,L}(h) }{\partial h_{w_k}\cdots \partial h_{w_1}}.
$$
\\

Proof.}  
We obtain the formula (\ref{kth}) with  $k$ times use of  integration by parts.
This implies  non negativity of all coefficients  of the Taylor series of the function $\gamma_i(v)$ around  any $v=v_0 \in [0,1)$.
Then,  $k$-th  derivatives are monotonically increasing in $v$.
From Taylor's theorem, there exists $v_1 \in (v_0,1)$ such that 
$$
\gamma_i'(v) =  \sum_{k=0}^{n-1}\frac{(v-v_0)^k}{k!} \gamma_i^{(k+1)} (v_0) +  \frac{(v-v_0)^n}{n!} \gamma_i^{(n+1)} (v_1).
$$   
Each term in this series  is bounded from the above by 
$$
\gamma_i'(1) =\frac{\beta^2 b^2 }{4|V_L|^2M} \sum_{w\in W_{L,M}} {\mathbb E} \langle \sigma_w \rangle_{b,i}^2
\leq \frac{\beta^2 b^2 }{4|V_L|}.
$$
This completes the proof. $\Box$\\

We define a term of the energy density with random field
\begin{eqnarray}
&&h^1_L:=\frac{1}{4|V_L| \sqrt{M}} \sum_{(x,t) \in W_{L,M}} h_{x,t} (1- \sigma_{x,t} \sigma_{x,t+1}), \\
&&h^3_L:=\frac{1}{2|V_L| \sqrt{M}} \sum_{w \in W_{L,M}} h_{x,t}  \sigma_{x,t}.
\end{eqnarray}
{\lemma  \label{delta2} 
For any $\beta b \neq 0$, 
we have 
\begin{equation}
{\mathbb E}\langle \delta { h^i_L}^2 \rangle_{b,i} \leq \frac{4\beta^2 b^2C}{   \sqrt{{|V_L|}}}
+\frac{C'}{|V_L| },
\label{a2}
\end{equation}
where $C$ and $C'$ are positive constant independent of $L$\\

Proof.}
For $i=3$ integration by parts and Lemma  \ref{delta}, \ref{FKGineq}  give
\begin{eqnarray}
{\mathbb E} \langle {\delta h^3_L}^2 \rangle_{b,3} 
&=&\frac{1}{4|V_L|^2 M} \sum_{x,y \in W_{L,M}}{\mathbb E} h_{x} h_y \langle  \sigma_{x} ; \sigma_y \rangle_{b,3} \nonumber \\
&=&\frac{1}{4|V_L|^2 M} \Big[\sum_{x,y \in W_{L,M}}{\mathbb E} \frac{\partial^2}{ \partial h_{x}\partial  h_y} 
\langle  \sigma_{x} ; \sigma_y \rangle_{b,3} +
\sum_{x\in W_{L,M}}{\mathbb E}\langle  \sigma_{x} ; \sigma_x \rangle_{b.c,3}  \Big]
\nonumber \\
&=&\frac{1}{4|V_L|^2 M} \Big[\sum_{x,y \in W_{L,M}}\frac{\beta^2b^2}{M}{\mathbb E} 
\langle  \sigma_{x} ; \sigma_y \rangle_{b,3} (\langle  \sigma_{x}  \sigma_y \rangle_{b,3}-3\langle  \sigma_{x} \rangle \langle \sigma_y \rangle_{b,3})+
\sum_{x\in W_{L,M}}{\mathbb E}\langle  \sigma_{x} ; \sigma_x \rangle_{b.c,3}  \Big]
\nonumber \\
&\leq&\frac{1}{|V_L|^2 M} \Big[\sum_{x,y \in W_{L,M}}\frac{\beta^2b^2}{M}{\mathbb E} 
\langle  \sigma_{x} ; \sigma_y \rangle_{b,3}+
\sum_{x\in W_{L,M}}{\mathbb E}(1-\langle  \sigma_x \rangle_{b.c,3}^2)  \Big]
\nonumber \\
&\leq &  \lim_{M\rightarrow \infty} 4\beta^2b^2 {\mathbb E}  \langle { \delta \mu^3_L}^2 \rangle_{b,1}+\frac{1}{|V_L| }
   \leq \frac{4\beta b^2C}{J_3\sqrt{[V_L|}} +\frac{1}{|V_L| }.\\
\end{eqnarray}
The bound for $i=1$ is obtained in the same way. $\Box$\\

{\lemma \label{limit} For almost all $b \in {\mathbb R}$,  we have
\begin{eqnarray}
&&\frac{\partial p_{i}}{\partial b} =
 \lim_{L \rightarrow \infty } \lim_{M \rightarrow \infty } \beta  {\mathbb E}\langle h^i_L  \rangle_{b,i} = \lim_{L \rightarrow \infty } \lim_{M \rightarrow \infty } \frac{\beta^2 b}{4}(1-{\mathbb E} \langle 
\rho_{1,2} ^i\rangle_{b,i}), \label{limit1}
\end{eqnarray}
for
$
\displaystyle{p_i(b):=\lim_{L \rightarrow \infty } \lim_{M \rightarrow \infty } p_{i,L}(b)},
$
and for $b\neq 0$, 
\begin{equation}
\lim_{L \rightarrow \infty } \lim_{M \rightarrow \infty } {\mathbb E} \langle |\Delta h_L^i |\rangle_{b,i} =0. \label{limit3}
\end{equation}

\noindent Proof.}  This can be shown in the standard convexity argument to obtain the Ghirlanda-Guerra identities in classical and quantum systems
\cite{C1,CG2,CG3,I,I2,Pn,T}. Note that $\psi_{i,L}$, $p_{i,L}$ and $p_{i}$ are convex  functions of $b$ and $c$. 
To show the first equality (\ref{limit1}), regard $p_{i,L}$ $p_i$ and $\psi_{i,L}$ as functions of $b$ for lighter notation. 
By Lemma \ref{1}, we have
$$
{\mathbb E}\psi_{i,L}(b)^2-p_{i,L}(b)^2 \leq \frac{C}{|V_L|}, 
$$
where $C$ is a positive number independent of $L$. 
Define the following functions 
\begin{eqnarray}
w_L(\epsilon) &:=& \frac{1}{\epsilon}[|\psi_{i,L}(b+\epsilon )-p_{i,L}(b+\epsilon)|+|\psi_{i,L}(b- \epsilon)-p_{i,L}(b-\epsilon)|
+|\psi_{i,L}(b )-p_{i,L}(b)| ]\nonumber \\
e_L(\epsilon )&:=&\frac{1}{\epsilon}[|p_{i,L}(b+\epsilon )-p_{i}(b+\epsilon)|+|p_{i,L}(b- \epsilon)-p_{i}(b-\epsilon)|
+|p_{i,L}(b )-p_{i}(b)|],\nonumber
\end{eqnarray}
for $\epsilon > 0$.
Note that the assumption on $\psi_{i,L}$  gives 
\begin{equation}
{\mathbb E}w_L(\epsilon) \leq \frac{3}{\epsilon} \sqrt{\frac{C}{|V_L|}},
\end{equation}
for any $\epsilon > 0$. 
Since $\psi_{i,L}$, $p_{i,L}$ and $p_i$ are convex functions of $b$, we have
\begin{eqnarray} 
&&\frac{\partial \psi_{i,L}}{\partial b}(b) - \frac{\partial  p_{i}}{\partial b}(b)
\leq \frac{1}{\epsilon} [\psi_{i,L}(b+\epsilon)-\psi_{i,L}(b)]- \frac{\partial  p_{i}}{\partial b}\nonumber \\
&&\leq \frac{1}{\epsilon} [\psi_{i,L}(b+\epsilon)-p_{i,L}(b+\epsilon)+p_{i,L}(b+\epsilon)-p_{i,L}(b)
+p_{i,L}(b)-\psi_{i,L}(b) \nonumber \\
&& -p_i(b+\epsilon) +p_i(b+\epsilon)+p_i(b)-p_i(b) ]- \frac{\partial  p_{i}}{\partial b}(b) \nonumber \\
&&\leq \frac{1}{\epsilon} [ |\psi_{i,L}(b+\epsilon)-p_{i,L}(b+\epsilon)|
+|p_{i,L}(b)-\psi_{i,L}(b)| +|p_{i,L}(b+\epsilon)-p_{i}(b+\epsilon)|\nonumber \\ 
&&+|p_{i,L}(b)-p_{i}(b)| ]+\frac{1}{\epsilon}[ p_{i}(b+\epsilon)-p_{i}(b)] - \frac{\partial  p_{i}}{\partial b}(b) \nonumber \\
&&\leq w_L(\epsilon) +e_L(\epsilon)
+  \frac{\partial p_{i}}{\partial b}(b+\epsilon) - \frac{\partial  p_{i}}{\partial b}(b). \nonumber
\end{eqnarray}  
As in the same calculation, we have
\begin{eqnarray} 
&&\frac{\partial \psi_{i,L}}{\partial b}(b) - \frac{\partial  p_{i}}{\partial b}(b) 
\geq \frac{1}{\epsilon}[\psi_{i,L}(b)-\psi_{i,L}(b-\epsilon)] - \frac{\partial  p_{i}}{\partial b}(b) \nonumber \\&&
\geq -w_L(\epsilon) -e_L(\epsilon)+ \frac{\partial p_{i}}{\partial b}(b-\epsilon)- \frac{\partial  p_{i}}{\partial b}(b) . \nonumber 
\end{eqnarray}  
Then, 
\begin{eqnarray} 
{\mathbb E}\Big|\frac{\partial \psi_{i,L}}{\partial b}(b) - \frac{\partial p_{i}}{\partial b}(b)\Big| \leq \frac{3}{\epsilon} \sqrt{\frac{ C}{|V_L|}} +e_L(\epsilon)+  \frac{\partial p_{i}}{\partial b}(b+\epsilon) -  \frac{\partial p_{i}}{\partial b}(b-\epsilon).\nonumber
\end{eqnarray}  
Convergence of $p_{i,L}$ in  the infinite volume limit implies 
\begin{eqnarray}
&&\lim_{L \rightarrow \infty }\lim_{M \rightarrow \infty } {\mathbb E}\Big| \beta \langle h^i_L  \rangle_{b,i} - \frac{\partial p_{i}}{\partial b}(b)\Big|
\leq  \frac{\partial p_{i}}{\partial b}(b+\epsilon) -  \frac{\partial p_{i}}{\partial b}(b-\epsilon), \nonumber
\end{eqnarray}
The right hand side vanishes, since the convex function$p_{i}(b)$ is continuously 
differentiable almost everywhere and $\epsilon >0$ is arbitrary.
  Jensen's inequality gives 
\begin{equation}
\lim_{L\rightarrow \infty} \lim_{M \rightarrow \infty }\Big| {\mathbb E}\beta  \langle h^i_L  \rangle_{b,i} - \frac{\partial p_{i}}{\partial b}(b)\Big|=0,
\label{limit3}
\end{equation}
for almost all $b$.  This leads the first equality (\ref{limit1}).  
The equality (\ref{limit3}) implies also
$$
\lim_{L \rightarrow \infty } \lim_{M \rightarrow \infty } {\mathbb E} |\langle  \Delta h^i_L  \rangle_{b,i}| =0.
$$
This and  Lemma \ref{delta2} enable us to obtain
$$
\lim_{L \rightarrow \infty } \lim_{M \rightarrow \infty } {\mathbb E} \langle |\Delta h^i_L|\rangle_{b,i} =0,
$$
since
 $$
{\mathbb E} \langle |\Delta h^i_L | \rangle_{b,i} =
 {\mathbb E} \langle |\delta h^i_L+ \langle \Delta h^i_L\rangle_{b,i} | \rangle_{b,i}
\leq {\mathbb E} \langle |\delta h^i_L | \rangle_{b,i} + {\mathbb E} |
\langle \Delta h^i_L\rangle_{b,i} | \leq 
\sqrt{{\mathbb E} \langle {\delta h^i_L} ^2 \rangle_{b,i} }+ {\mathbb E} |
\langle \Delta h^i_L\rangle_{b,i} | .
$$
Therefore
the identities are obtained from the above as in the random field Ising model \cite{C2}.
$\Box$\\

Note that Lemma \ref{limit} implies the existence of 
$\displaystyle{\lim_{L \rightarrow \infty } \lim_{M \rightarrow \infty }{\mathbb E} \langle \rho^i_{1,2}\rangle_{b,i}}$ for $b\neq 0$.\\


{\lemma \label{GGid} 
Let $f:{\cal S}_W^n \rightarrow {\mathbb R}$ be a bounded function of $n$ replicated spin configurations. 
 The Gibbs and sample expectations of $f$ and spin overlap  in the model defined by the Hamiltonian (\ref{hamilu}), 
 satisfy the following identity  for almost all $b \in {\mathbb R}$
\begin{equation}
\lim_{L \rightarrow \infty } \lim_{M \rightarrow \infty }[ {\mathbb E} \langle  f \rho_{1,n+1}^i \rangle_{b,i} - \frac{1}{n}{\mathbb E} \langle f \rangle_{b,i}   {\mathbb E} \langle \rho_{1,2}^i \rangle _{b,c,i}- \frac{1}{n}\sum_{\alpha=2} ^n {\mathbb E}  \langle 
 f  \rho^i_{1,\alpha}\rangle_{b,i} ]=0,
\end{equation}
which provides the Ghirlanda-Guerra identities. \\

Proof.} 
From the identity (\ref{limit3}) in Lemma \ref{limit},
$$
\lim_{L \rightarrow \infty } \lim_{M \rightarrow \infty } {\mathbb E} \langle \Delta h^i_Lf \rangle_{b,i} =0.
$$
Calculating the right hand side gives the identity. $\Box$ \\

{\lemma \label{cont} 
For almost all constant field $c \in {\mathbb R}$, the expectation of the overlap in the infinite volume limit  is continuous at $b=0$
\begin{eqnarray}
\lim_{b\rightarrow 0} \lim_{L\rightarrow \infty}  \lim_{M\rightarrow \infty}{\mathbb E} \langle \rho_{1,2}^i \rangle_{b,i} =\lim_{L\rightarrow \infty}  \lim_{M\rightarrow \infty}{\mathbb E} \langle \rho_{1,2}^i \rangle_{0,i}, \label{1st} \\
\lim_{b\rightarrow 0} \lim_{L\rightarrow \infty}  \lim_{M\rightarrow \infty}{\mathbb E} \langle \rho_{1,2}^i \rangle_{b,i}^2 =\lim_{L\rightarrow \infty}  \lim_{M\rightarrow \infty}{\mathbb E} \langle \rho_{1,2}^i \rangle_{0,i}^2. \label{2nd}
\end{eqnarray}
Proof. }
Evaluate the following  partial derivative 
\begin{eqnarray}
&&
\Big| \frac{\partial}{\partial b}{\mathbb E}  \langle \rho^3_{1,2} \rangle_{b,3} \Big| \nonumber \\
&&=\Big|  \frac{\beta^2 b}{8|V_L| M^2} \sum_{x,y\in W_{L,M}} {\mathbb E}  \langle \sigma_x;\sigma_y \rangle_{b,3}
( \langle \sigma_x\sigma_y \rangle_{b,3}  -3\langle \sigma_x  \rangle_{b,3}  \langle \sigma_y \rangle_{b,3})\Big|\nonumber \\
&&\leq \frac{\beta^2b}{8|V_L| M^2} \sum_{x,y\in W_{L,M}} {\mathbb E}|  \langle \sigma_x;\sigma_y \rangle_{b,3}||
\langle \sigma_x\sigma_y \rangle_{b,3}  -3\langle \sigma_x  \rangle_{b,3}  \langle \sigma_y \rangle_{b,3} | \nonumber \\
&&\leq\frac{\beta^2b}{2|V_L| M^2} \sum_{x,y\in W_{L,M}} {\mathbb E}|\langle \sigma_x;\sigma_y \rangle_{b,3}|  
\leq\frac{\beta^2b}{2|V_L| M^2} \sum_{x,y\in W_{L,M}} {\mathbb E}  \langle \sigma_x;\sigma_y \rangle_{b,3}  
\nonumber \\&&
\leq 
 \frac{\beta^2b}{2|V_L| M^2} \sum_{x,y\in W_{L,M}} {\mathbb E}  \langle \sigma_x;\sigma_y \rangle_{b,3} 
\leq2 \beta b\frac{\partial}{\partial c}{\mathbb E}  \langle \mu^3_L \rangle_{b,3} 
\end{eqnarray}
The FKG inequality has been used. This bound enables us to evaluate the following integral 
\begin{eqnarray}
&& \int_{c_1} ^{c_2} dc | \lim_{L\rightarrow\infty} \lim_{M\rightarrow \infty} [
{\mathbb E} \langle \rho_{1,2}^3 \rangle_{b,3}-{\mathbb E} \langle \rho_{1,2}^3 \rangle_{0,3}] |
= 
 \lim_{L\rightarrow\infty}   \int_{c_1} ^{c_2}dc \Big| \int_0^{b} db' \frac{\partial }{\partial b'} 
  \lim_{M\rightarrow \infty}  {\mathbb E} \langle \rho_{1,2}^3 \rangle_{b',3}\Big| \nonumber \\
&&\leq   \lim_{L\rightarrow\infty} 2 \beta \int_{c_1} ^{c_2}dc \int_0^{b} db'
b' \frac{\partial }{\partial c}  \lim_{M\rightarrow \infty}  {\mathbb E} \langle \mu_{1,2}^3 \rangle_{b',3} \nonumber \\
&&=2\beta  \int_{0} ^{b}db' b'  \lim_{L\rightarrow\infty}  \lim_{M\rightarrow \infty} 
 [{\mathbb E} \langle \mu_L^3 \rangle_{b',3,c=c_2}-{\mathbb E} \langle \mu_L^3 \rangle_{b',3,c=c_1}].
\end{eqnarray} 
The boundedness of ${\mathbb E} \langle \mu_L^3 \rangle_{b,3}$ gives
the limit 
\begin{equation}
 \int_{c_1} ^{c_2} dc |\lim_{b\rightarrow 0} \lim_{L\rightarrow\infty} \lim_{M\rightarrow \infty} [
{\mathbb E} \langle \rho_{1,2}^3 \rangle_{b,3}-{\mathbb E} \langle \rho_{1,2}^3 \rangle_{0,3}] |
= 0
\end{equation}
for arbitrary $c_1, c_2 \in {\mathbb R}$. Therefore, the integrand in the left hand side vanishes for almost all $c$, and this implies the 
first equality (\ref{1st})  for $i=3$. 

For $i=1$,   evaluate the partial derivative
\begin{eqnarray}
&&
\Big| \frac{\partial}{\partial b}{\mathbb E}  \langle \rho^1_{1,2} \rangle_{b,1} \Big| \nonumber \\
&&=\Big| \frac{\beta^2 b}{64|V_L| M^2} \sum_{x,y\in V_L,s,t\in T_M} {\mathbb E}  \langle \sigma_{x,s} \sigma_{x,s+1};\sigma_{y,t} \sigma_{y,t+1} \rangle_{b,1}( \langle \sigma_{x,s} \sigma_{x,s+1};\sigma_{y,t} \sigma_{y,t+1} \rangle_{b,1} 
\nonumber \\&&-3\langle \sigma_{x,s} \sigma_{x,s+1}\rangle_{b,1}\langle \sigma_{y,t} \sigma_{y,t+1} \rangle_{b,1}+2)\Big|
\nonumber \\
&&\leq \frac{3\beta^2b}{32|V_L| M^2} \sum_{x,y\in V_L,s,t\in T_M} {\mathbb E}|  \langle \sigma_{x,s} \sigma_{x,s+1};\sigma_{y,t} \sigma_{y,t+1} \rangle_{b,1}|
\nonumber \\
&&\leq \frac{3\beta^2b}{32|V_L| M^2} \sum_{x,y\in V_L,s,t\in T_M}{\mathbb E}\langle (\sigma_{x,s} +\sigma_{x,s+1});(\sigma_{y,t} +\sigma_{y,t+1} ) \rangle_{b,1}
\nonumber \\&&
\leq\frac{3\beta^2b}{8|V_L| M^2} \sum_{x,y\in W_{L,M}} {\mathbb E}  \langle \sigma_x;\sigma_y \rangle_{b,1}  
\nonumber \\&&
 \leq \frac{3\beta b}{2}\frac{\partial}{\partial c}{\mathbb E}  \langle \mu^3_L \rangle_{b,1} 
\end{eqnarray}
The inequality (\ref{bondcorr}) in Lemma \ref{s1} has been used. This bound and the same argument as for $i=3$
give  the first equality (\ref{1st}) for $i=1$.

To show the second equality (\ref{2nd}), 
the following representation obtained by the FKG inequality 
is useful
\begin{eqnarray}
\frac{\partial }{\partial b} {\mathbb E}\langle \rho^3_{1,2} \rangle_{b,3}^2  
&&=  \frac{\partial }{\partial b} {\mathbb E}\Big( \frac{1}{4|V_L|M} \sum_{w \in W_{L,M}} \langle \sigma_w  \rangle_{b,3}^2  \Big)^2  
\nonumber \\
&&=
\frac{\beta^2 b}{16|V_L|^2M^2} \sum_{x,y,z \in W_{L,M}}{\mathbb E}  \langle \sigma_y; \sigma_z\rangle_{b,3} \langle \sigma_x\rangle_{b,3} 
(2\langle \sigma_x; \sigma_z\rangle_{b,3} \langle \sigma_y\rangle_{b,3} \nonumber \\
&&-2 \langle \sigma_x \rangle_{b,3} \langle \sigma_y\rangle_{b,3} \langle \sigma_z\rangle_{b,3}+ \langle \sigma_x\rangle_{b,3}  \langle \sigma_y; \sigma_z\rangle_{b,3}) \nonumber  \\
&&\leq \frac{ \beta^2 b}{2|V_L|M} \sum_{x,y,z \in W_{L,M}}{\mathbb E}  \langle \sigma_y; \sigma_z\rangle_{b,3},
\nonumber \\ &&=2\beta b \frac{\partial }{\partial c}  {\mathbb E}  \langle \mu_L^3\rangle_{b,3},
\end{eqnarray}
This bound  and the boundedness of $ {\mathbb E}  \langle \mu_L^3\rangle_{b,3}$
enable us to prove the
 second equality (\ref{2nd}) as well as the first one (\ref{1st}).  The second equality (\ref{2nd}) for $i=1$ is proved 
 by  showing the bound
 $$\frac{\partial }{\partial b} {\mathbb E}\langle \rho^1_{1,2} \rangle_{b,1}^2\leq  \frac{5\beta b}{8} \frac{\partial }{\partial c}{\mathbb E}  \langle \mu_L^3\rangle_{b,1}.
 $$ 
This bound and the boundedness of $ {\mathbb E}  \langle \mu_L^3\rangle_{b,1}$
enable us to prove the
 second equality (\ref{2nd}) for $i=1$, and this 
 completes the proof.
$\Box$ \\

\paragraph{Proof of }Theorem \ref{MT}\\
 Since    $S_x^2 | E \rangle $ is orthogonal to $| E \rangle $  for  an arbitrary 
eigenstate $| E \rangle $ of the Hamiltonian, we obtain 
$\langle S_x^2\rangle =0$ and $\langle S_x^2 S_y^2 \rangle =\delta_{x,y}/4$. 
These imply
$${\mathbb E} \langle R^2_{1,2}\rangle=0, \ \ \ {\mathbb E} \langle{ R^2_{1,2}}^2 \rangle= \frac{1}{16|V_L|},
$$
then Theorem \ref{MT} is valid  trivially for $R_{1,2}^2$. Therefore, we consider $R_{\alpha,\beta}^i$ for $i=1,3$.
Since 
$\displaystyle{\lim_{L \rightarrow \infty } \lim_{M \rightarrow \infty }{\mathbb E}  
\langle \rho_{1,2}^i \rangle_{b,i}}
$
exists by Lemma \ref{limit},   this limit exists   
 also for $b=0$ by Lemma \ref{cont}. 
$$
\lim_{L \rightarrow \infty } \langle R^i_{1,2} \rangle=\lim_{L \rightarrow \infty } \lim_{M \rightarrow \infty }{\mathbb E} 
\langle \rho_{1,2}^i \rangle_{0,i}=\lim_{b \rightarrow 0 }\lim_{L \rightarrow \infty } \lim_{M \rightarrow \infty }{\mathbb E}  
\langle \rho_{1,2}^i \rangle_{b,i}.
$$
First, we use the Ghirlanda-Guerra identities  for $b\neq0.$

 For $n=2$ and $f=\rho^i_{1,2}$, the identity in Lemma \ref{GGid} 
 \begin{equation}
\lim_{L \rightarrow \infty }\lim_{M \rightarrow \infty }[ 2{\mathbb E} \langle  \rho^i_{1,2} \rho^i_{1,3}\rangle_{b,i} - 
({\mathbb E} \langle \rho^i_{1,2} \rangle_{b,i} )^2 - {\mathbb E}  \langle{ \rho^i_{1,2}}^2\rangle_{b,i}  ]=0.
 \label{GG2}
\end{equation}
For $n=3$ and $f=\rho^i_{2,3}$, the identity in Lemma \ref{GGid} gives
 \begin{equation}
\lim_{L \rightarrow \infty }\lim_{M \rightarrow \infty }[3 {\mathbb E} \langle  \rho^i_{2,3} \rho^i_{1,4}\rangle_{b,i} - 
({\mathbb E} \langle \rho^i_{1,2} \rangle_{b,i} )^2 - {\mathbb E}  \langle \rho^i_{2,3} \rho^i_{1,2}\rangle_{b,i} - {\mathbb E}  \langle \rho^i_{2,3} \rho^i_{1,3}\rangle_{b,i}  ]=0.
 \label{GG3}
\end{equation}
These two identities and 
 $ \langle \rho^i_{2,3} \rho^i_{1,4}\rangle_{b,i}= \langle  {\rho^i_{1,2}}^2 \rangle_{b,i} $ and
$ \langle \rho^i_{1,2} \rho^i_{1,3}\rangle_{b,i}= \langle \rho_{2,3} \rho^i_{1,2}\rangle_{b,i} = \langle \rho^i_{2,3} \rho^i_{1,3}\rangle_{b,i}$
in the replica symmetric Gibbs state  imply
$$
2 \lim_{L \rightarrow \infty }\lim_{M \rightarrow \infty }[ {\mathbb E} \langle  \rho^i_{1,2}\rangle_{b,i}^2- ({\mathbb E} \langle \rho^i_{1,2} \rangle_{b,i} )^2] 
= \lim_{L \rightarrow \infty }\lim_{M \rightarrow \infty }[ {\mathbb E} \langle  {\rho^i_{1,2}}^2\rangle_{b,i}- {\mathbb E} \langle  \rho^i_{1,2}\rangle_{b,i}^2]
$$
Since the right hand side vanishes in the above for any $b$ because of Lemma \ref{MT0},
the left hand side vanishes for almost all $b\neq0$.  This fact and Lemma \ref{cont}  imply that
the left hand side vanishes also for $b=0$. Then, we have
\begin{eqnarray}
&&\lim_{L \rightarrow \infty} [{\mathbb E} ( R^i_{1,2}, R^i_{1,2})_{\rm D} - ({\mathbb E} \langle {R^i_{1,2}}\rangle)^2]=
\lim_{b \rightarrow 0} \lim_{L \rightarrow \infty}  \lim_{M \rightarrow \infty }[ {\mathbb E} \langle  {\rho^i_{1,2}}^2 \rangle_{b,i}- ({\mathbb E} \langle \rho^i_{1,2} \rangle_{b,i} )^2]  \nonumber \\
&&=\lim_{b \rightarrow 0} \lim_{L \rightarrow \infty}  \lim_{M \rightarrow \infty }[ {\mathbb E} \langle  {\rho^i_{1,2}}^2 \rangle_{b,i}-{\mathbb E} \langle  \rho^i_{1,2}\rangle_{b,i}^2+{\mathbb E} \langle  \rho^i_{1,2}\rangle_{b,i}^2- ({\mathbb E} \langle \rho^i_{1,2} \rangle_{b,i} )^2] =0.
\nonumber
\end{eqnarray}
Harris'  inequality of the Bogolyubov type
between the Duhamel product and the Gibbs expectation of the square of arbitrary self-adjoint operator $O$ \cite{H}
\begin{equation}
( O,  O)_{\rm D} \leq \langle {O}^2\rangle \leq ( O,  O)_{\rm D}  +\frac{\beta}{12} \langle[ O, [H,O]] \rangle,
\label{harris}
\end{equation}
enables us to obtain  
$$
\lim_{L \rightarrow \infty} {\mathbb E} ( R^i_{1,2}, R^i_{1,2})_{\rm D} =\lim_{L \rightarrow \infty}  {\mathbb E} \langle {R^i_{1,2}}^2\rangle. 
$$
Therefore
\begin{equation}
\lim_{L \rightarrow \infty}
 [  {\mathbb E} \langle {R^i_{1,2}}^2 \rangle-({\mathbb E} \langle {R^i_{1,2}}\rangle)^2]=0.
\label{1112}
\end{equation}
 This completes
the proof of Theorem \ref{MT}. $\Box$\\

 
{\bf Acknowledgments}\\

 It is pleasure  to thank R. M. Woloshyn  for reading the manuscript and helpful suggestions.  
 I am grateful to M. Aoyagi for discussions in early stage of this work. I would like to thank the anonymous referees for essential comments.


\begin{thebibliography}{13}
\bibitem{AC} Aizenman, M.,  Contucci, P. :{ On the stability of quenched state in mean-field spin glass models}. J. Stat. Phys. \textbf{92}, 765-783(1997)

\bibitem{AGL} Aizenman, M., Greenblatt,R.L.,  Lebowitz, J. L. :{Proof of rounding by quenched disorder of first order transitions in low-dimensional quantum systems} J. Math. Phys. {\textbf 53} 10.1063, (2012)



\bibitem{C2} Chatterjee, S. : { Absence of replica symmetry breaking in the random field Ising model}. Commun. Math .Phys. \textbf{337}, 93-102(2015)

\bibitem{C1} Chatterjee,S.:{ The Ghirlanda-Guerra identities without averaging}. preprint, arXiv:0911.4520 (2009).

\bibitem{C} Chatterjee, S. :{ Disorder chaos and multiple valleys in spin glasses}. preprint, arXiv:0907.3381 (2009).  


\bibitem{CG2} Contucci, P., Giardin\`a, C. :{ The Ghirlanda-Guerra identities}. J. Stat. Phys. \textbf{ 126}, 917-931,(2007)

\bibitem{CG3} Contucci, P., Giardin\`a, C. :{ Perspectives on spin glasses.} Cambridge university press, 2012.

\bibitem{CGP} Contucci, P.,  Giardin\`a, C., Pul\'e, J. :{ The infinite volume limit for finite dimensional classical and quantum disordered systems}. Rev.  Math. Phys. 
\textbf{ 16}, 629-638, (2004)

\bibitem{CK} Campanino, M., Klein, A. :{Decay of Two-Point Functions for (d + 1)-Dimensional Percolation, Ising and Potts Models with d-Dimensional Disorder}. Commun. Math .Phys. \textbf{135}, 483-497(1991)

\bibitem{CKP} Campanino, M., Klein, A., Pelez, J. F.,  :{Localization in the Ground State of the Ising Model with a Random Transverse Field}. Commun. Math. Phys. 135, 499-515 (1991)

\bibitem{CL} Contucci, P., Lebowitz, J. L. :{ Correlation inequalities for quantum spin systems with quenched centered disorder}. J. Math. Phys. \textbf{ 51}, 023302-1 -6 (2010)

\bibitem{Cr} Crawford, N. :{ Thermodynamics and universality for mean field quantum spin glasses.} Commun. Math. Phys.
\textbf{274}, 821-839(2007) 



 \bibitem{EA}  Edwards,S. F., Anderson, P. W. :{ Theory of spin glasses} J. Phys. F: Metal Phys. \textbf{5}, 965-974(1975)


\bibitem{FKG}  Fortuin,C. M., Kasteleyn P. W.,  Ginibre, J.:{ Correlation inequalities on some partially ordered sets}.Commun. Math. Phys. \textbf{22}, 89-103(1971). 



 \bibitem{GG} Ghirlanda, S., Guerra, F. :{ General properties of overlap probability distributions in disordered spin systems. Towards Parisi ultrametricity}.
 J. Phys. A\textbf{31}, 9149-9155(1998)

\bibitem{GUW} Goldschmidt, C., Ueltschi, D., Windridge, P:{Quantum Heisenberg models and their probabilistic representations} 
Entropy and the quantum II, Contemp. Math. {\textbf 562} 177-224, (2011) 


 





\bibitem{H}  Harris, A.B. :{Bounds for certain thermodynamic averages}J. Math. Phys. 8 1044-1045.(1967) 

 


\bibitem{I} Itoi, C. :{General properties of overlap operators in disordered quantum spin systems} J. Stat. phys. {\textbf 163} 1339-1349 (2016) 

 
\bibitem{I2} Itoi, C. :{Universal nature  of replica symmetry breaking 
 in quantum systems with Gaussian disorder} J. Stat. phys. {\textbf 163} 1339-1349 (2016) 
 


\bibitem{NS1} H. Nishimori and D. Sherrington, AIP Conference Proceedings 553, 67 (2001)

\bibitem{N} H. Nishimori, ``Statistical Physics of Spin Glasses and Information Processing: An Introduction"
Oxford university press (2001)

\bibitem{Pn} Panchenko, D. :{ The Ghirlanda-Guerra identities for mixed $p$-spin glass model}. Compt. Read. Math. \textbf{ 348}, 189-192(2010).
 


\bibitem{Pr} Parisi, G. :{A sequence of approximate solutions to the S-K model for spin glasses}. J. Phys. A \textbf{13}, L-115 (1980)

\bibitem{S} Seiler, E., Simon, B. :{ Nelson's symmetry and all that in Yukawa and $(\phi^4)_3$ theories}. Ann.  Phys. \textbf{ 97}, 470-518, (1976)

\bibitem{SK} Sherrington, S., Kirkpatrick, S :{ Solvable model of spin glass}. Phys. Rev. Lett. \textbf{ 35},  1792-1796, (1975). 
 
\bibitem{Suzuki}  M. Suzuki, :{Relationship between d-dimensional quantal spin systems and (d+1)-dimensional Ising systems}. Prog. Theor. Phys. {\textbf 56}, 1454-1468 (1976)


\bibitem{T2}  Talagrand,  M. :{ The Parisi formula}. Ann. Math. \textbf{ 163}, 221-263 (2006).

\bibitem{T} Talagrand, M. :{ Mean field models for spin glasses}. Springer, Berlin (2011).
	 






 


 
 

\end{thebibliography}
\end{document}